\documentclass[a4paper]{article}
\usepackage[square,sort&compress,numbers]{natbib}

\usepackage{amsmath,amssymb} 
\usepackage{physics}
\usepackage{relsize}
\usepackage{booktabs}
\usepackage{soul} 
\usepackage{graphicx}
\usepackage{xcolor}

\usepackage{fancyhdr}

\usepackage[colorlinks=true,allcolors=blue]{hyperref}

\renewcommand{\i}{{\rm i}}

\newcommand{\Hc}{{\rm H.c.}}
\renewcommand{\title}[1]{{\exhyphenpenalty=10000\hyphenpenalty=10000 
		\fontsize{18}{21}\selectfont\noindent\raggedright
		\textsf{#1}\par}\suppressfloats[t]}

\renewcommand{\author}[1]{{\vspace{5mm}%
		\fontsize{10}{12}
		\raggedright \if@anonymous Author list removed for anonymity \else #1 \fi
		\vspace{3mm}}}
\newcommand{\orcid}[1]{\href{https://orcid.org/#1}{\includegraphics[width=8pt]{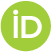}}}
\newcommand{\keywords}[1]{{\fontsize{8}{10}\selectfont
		\raggedright {\bfseries Keywords:} #1}
}

\renewenvironment{abstract}{%
	\vspace{16pt plus3pt minus3pt}
	{\color{gray}\hrule} \ \\
	\noindent \fontsize{11}{12}\selectfont {\bfseries Abstract}\\
	\rm\ignorespaces \raggedright}{\vspace{3mm} {\color{gray}\hrule}}	
	
\renewenvironment{abstract}{%
	\vspace{16pt plus3pt minus3pt}
	{\color{gray}\hrule} \ \\
	\noindent \fontsize{11}{12}\selectfont {\bfseries Abstract}\\
	\rm\ignorespaces \raggedright}{\vspace{3mm} {\color{gray}\hrule}}
\begin{document}

\title{\textbf{One-step preparation of 3D Bell and 3D GHZ states with Rydberg atoms}}

\author{Jiping Wang$^{2,*}$\orcid{0009-0004-6742-602X}  and  Huapeng Liu$^{1,*}$}


  \begin{abstract}
  Three-dimensional Bell states and GHZ states serve as representative examples of high-dimensional entangled states. In this paper, we propose a scheme for generating three-dimensional Bell and GHZ entangled states using Rydberg atoms. By leveraging Rydberg-mediated interactions and introducing detuning, the system is effectively simplified into a chain-like configuration. To design effective couplings, we employ a centrosymmetric Gaussian distribution and optimize the relevant parameters. Furthermore, we take into account decoherence factors including atomic spontaneous emission, dephasing effects and random noise. Numerical simulations indicate that the proposed scheme can achieve high fidelity.

  \end{abstract}

\section{Introduction}\label{introduction}
Quantum entanglement is a fundamental resource for quantum computing and quantum communication, with broad application prospects in quantum information processing (QIP). Two-dimensional entanglement represents the most fundamental form of quantum entanglement and has been widely applied to a variety of  QIP tasks, including quantum computing\cite{i1}, quantum teleportation\cite{i2,hu2020experimental} and  quantum cryptography\cite{i3}. However, compared to their two-dimensional counterparts, high-dimensional entangled states offer greater advantages in both fundamental research and practical applications\cite{i7,cozzolino2019high}. It not only enable stronger violations of local realism\cite{i4}, but also significantly enhance the security of quantum cryptographic\cite{i5,i6}. In the recent past years, numerous schemes have been proposed for the generation of high-dimensional entangled states\cite{g1,g2,g3,g4,wang2020generation,shao2014stationary,li2011deterministic,zhao2024dissipative}. Among the various types of high-dimensional entangled states, the three-dimensional Bell states\cite{wang2020generation, shao2014stationary,li2011deterministic} and GHZ states\cite{zhang2011generation,zhao2024dissipative,g3} are particularly noteworthy. One of their maximally entangled forms can be expressed as
\begin{align}
&\ket{\mathrm{Bell}} = \frac{1}{\sqrt{3}}(\ket{00} + \ket{11} + \ket{22}), \cr
&\ket{\mathrm{GHZ}} = \frac{1}{\sqrt{3}}(\ket{000} + \ket{111} + \ket{222}).
\end{align}
They are both types of states exhibit stronger violations of generalized Bell inequalities~\cite{dada2011experimental}, making them valuable tools for testing the foundations of quantum mechanics and for implementing more robust and secure quantum information protocols.

On the other hand, neutral atoms with long-lived Rydberg states have shown great potential for the generation of entanglement. Rydberg states refer to highly excited atomic states characterized by {their long lifetimes, which can reach several hundred microseconds ($\mu$s) and even up to the millisecond range}\cite{holzl2024long} ,  far exceeding those of low-lying excited states\cite{theodosiou1984lifetimes,archimi2019measurements,mack2015all,holzl2024long}.{This extended stability makes them suitable for quantum information encoding}\cite{cohen2021quantum,saffman2010quantum,ji2020fast,zhao2017rydberg}. When an atom is excited to a Rydberg state, strong dipole–dipole or van der Waals interactions arise between Rydberg atoms. These interactions significantly shift the energy levels of nearby atoms in Rydberg states. As a result, when one atom is excited to a Rydberg state, other atoms within a certain distance (blockade radius) are prevented from being excited to Rydberg states, this phenomenon is referred to as the Rydberg blockade\cite{ryb1,ryb2,ryb3,ryb4,ryb5}. Under this effect, at most one atom can occupy the Rydberg state within the blockade region. Conversely, the Rydberg antiblockade effect is a quantum phenomenon that contrasts with the Rydberg blockade. In the standard Rydberg blockade mechanism, due to dipole–dipole or van der Waals interactions, when one atom is excited to a Rydberg state, its neighboring atoms cannot be excited to the same Rydberg state. However, under certain conditions, such as tuning the frequency of the laser field, multiple atoms can be simultaneously excited to Rydberg states, thereby overcoming the blockade limitation; this phenomenon is referred to as the Rydberg antiblockade\cite{rab1,rab2,rab3}. Due to the intriguing mechanisms of Rydberg atoms, numerous schemes for the preparation of high-dimensional entangled states have been proposed. For example, in 2018, Li and Shao \cite{li2018unconventional} proposed a method called unconventional Rydberg pumping and used it to prepare three-dimensional Bell states.  In 2020, Da et al. \cite{da2020efficient} employed the Rydberg blockade mechanism to prepare a three-particle singlet state. Wang et al. \cite{wang2020generation} utilized the Rydberg antiblockade effect to generate three-dimensional Bell states.  In 2024, Zhao et al. \cite{zhao2024dissipative} combined Rydberg atoms and dissipative processes to  prepare three-dimensional GHZ states.

Inspired by \cite{wang2025deterministic,wu2021resilient,zheng2020deterministic,wang2020generation,zhao2024dissipative}, we propose a protocol to realize three-dimensional Bell states and three-dimensional GHZ states with Rydberg atoms. First, we simplify the system dynamics. By introducing laser detuning, we derive the effective Hamiltonian for the model preparing three-dimensional Bell states, where the effective energy-level structure forms a chain-like five-level system. Similarly, for the GHZ state preparation model, we initially introduce detuning to obtain an effective eight-level system; subsequently, through amplitude modulation of the Rabi frequencies, the final effective system is reduced to a chain-like six-level structure. For the resulting effective chain-like energy levels, various quantum coherent control techniques can be implemented by designing appropriate coupling profiles. These include square pulses\cite{vogel2006quantum}, Lyapunov control\cite{kuang2008lyapunov,lc1,lc2}, adiabatic evolution\cite{ae2,ae1,ae3}, machine learning\cite{ml1,ml2,xue2021preparation}, and optimal control\cite{oc1,oc2,oc3,oc4}. In this work, we set the effective couplings to have a centrosymmetric Gaussian profile and employ gradient descent and genetic algorithms to optimize the amplitude, width, and phase of the pulses to achieve the desired target states. Furthermore, we use the designed pulses to simulate the original Hamiltonian while incorporating decoherence factors, including atomic spontaneous emission and dephasing, to evaluate their impact on fidelity. 
{Compared to the previous scheme, our scheme has following advantages: (1) our scheme enables the direct preparation of high-dimensional Bell and GHZ states within a single evolution process, thereby reducing the overall complexity. (2) The theoretical construction method of the scheme is highly extensible, as the preparation of three-dimensional Bell states and GHZ states follows a similar procedure. }

The structure of this paper is arranged as follows. In Section \ref{sec2}, we present the model for the preparation of Bell states and GHZ states, with a focus on the design of the Rabi frequencies. In Section \ref{sec3}, numerical simulations and feasibility analysis are carried out. The summary and discussion in Section \ref{sec4}.

\section{Physical Model and effective }\label{sec2}
\subsection{Three-dimensiona Bell state}
\begin{figure}[ht]
    \centering
    \includegraphics[width=0.5\linewidth]{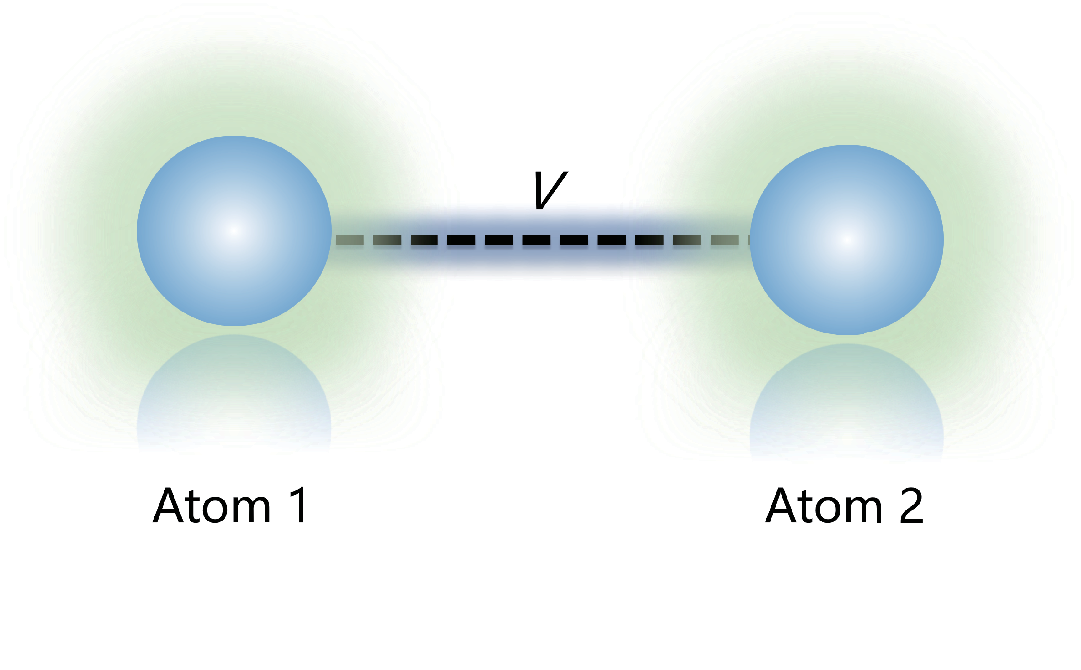}
    \caption{Two Rydberg atoms with the interaction strength $V$.}\label{twoa}
\end{figure}
\begin{figure}[ht]
    \centering
    \includegraphics[width=0.3\linewidth]{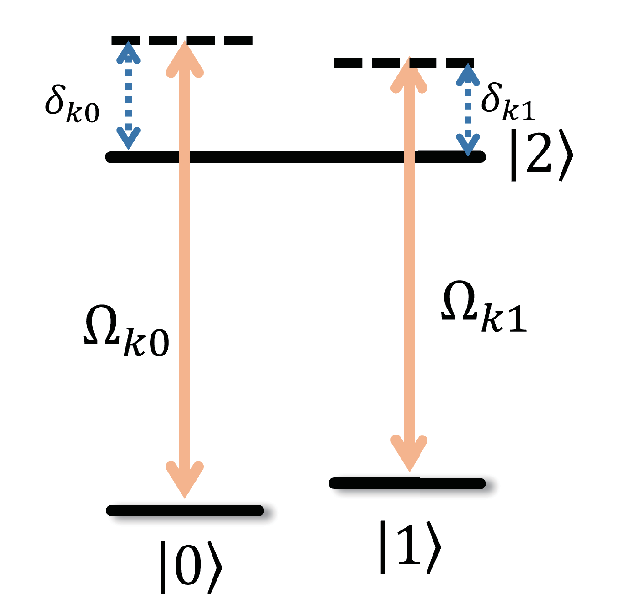}
    \caption{{Excitation scheme and relevant energy levels of atom $k$.}}\label{EL}
\end{figure}
We consider  two identical neutral atoms  are confined in optical tweezers as shown in Figure \ref{twoa}, and  atoms which has three energy levels as show in Figure \ref{EL}, the ground states $\ket{0}$, $\ket{1}$, and Rydberg state $\ket{2}$, The transition of $\ket{0}_k \leftrightarrow \ket{2}_k$,  $\ket{1}_k \leftrightarrow \ket{2}_k$ are driven by $\Omega_{k0}$ and $\Omega_{k1}$ with detunings $\delta_{k0}$, $\delta_{k1}$, respectively, where $k$ denotes the $k$th atom in the system.
Under the rotating wave approximation (RWA), then the Hamiltonian of the two-atom system in the interaction picture can be represented by
\begin{align}\label{eq1}
    H_2(t)=\sum_{k=1,2}\sum_{j=0,1}\left[{\Omega_{k,j}(t)e^{-\i \delta_{kj}t}\ket{2}_k\bra{j}+{\rm H.c.}}\right]+V\ket{22}\bra{22},
\end{align}
{the energy level shifts induced by the Rydberg states  are given by}
\begin{align}
    H_{2V}=V\ket{22}\bra{22}
\end{align}
{Next, we define a rotating frame with a unitary operator $\mathcal{R}={\rm exp} {(-\i H_{2V}t)}$, and move into the new picture through}
\begin{align}
	H_2'=\mathcal{R}^\dag H_2 \mathcal{R} -\i{\mathcal{R}^\dag} \Dot{\mathcal{R}}.
\end{align}
We can obtain
\begin{align}
	H_2'=&\Omega_{10}e^{-\i \delta_{10}t}\left(\ket{20}\bra{00}+\ket{21}\bra{01}+e^{\i Vt}\ket{22}\bra{02}\right) \cr
    &+\Omega_{11}e^{-\i \delta_{11}t}\left(\ket{20}\bra{10}+\ket{21}\bra{11}+e^{\i Vt}\ket{22}\bra{12}\right)\cr
    &+\Omega_{20}e^{-\i \delta_{20}t}\left(\ket{02}\bra{00}+\ket{12}\bra{10}+e^{\i Vt}\ket{22}\bra{20}\right)\cr
    &+\Omega_{21}e^{-\i \delta_{21}t}\left(\ket{02}\bra{01}+\ket{12}\bra{11}+e^{\i Vt}\ket{22}\bra{21}\right)+\Hc
\end{align}
when set the detuning of the Rabi frequencies as follows
\begin{align}    \delta_{10}=\delta_{21}=0,\delta_{20}=\delta_{11}=V,
\end{align}
And if initial state is $\ket{00}$ and through neglecting highly frequent oscillations under the condition $V\gg \Omega_{k,j}$, we can derive the effective Hamiltonian
\begin{align}
    H_{\rm eff}=\Omega_{1}\ket{20}\bra{00}+\Omega_{2}\ket{22}\bra{20}+\Omega_{3}\ket{22}\bra{12}+\Omega_{4}\ket{12}\bra{11}+{\rm H.c.}\label{belleff}
\end{align}
{where $\Omega_1=\Omega_{10}$, $\Omega_2=\Omega_{20}$, $\Omega_3=\Omega_{11}$, $\Omega_4=\Omega_{21}$.}
This effective Hamiltonian can be regarded as an chain-type five-level system.
 From a theoretical standpoint, it is feasible to design appropriate pulse sequences to generate the desired quantum states, and we can realize arbitrary superpositions of basis states within this subspace by tailoring the shape of Rabi frequencies  or  square pulses with specific proportional relationships.

From an experimental perspective, smooth-amplitude pluser can help mitigate off-resonant coupling to states not included in the idealized model, while also providing finite laser pulse bandwidth, thereby reducing the experimental complexity associated with rapid switching.
In the scheme for preparing the three-dimensional Bell state, the Rabi frequency is shaped as a symmetric Gaussian function, its amplitude vanishing at both the initial time and the final time $T$\cite{li2024high}.
\begin{align}
	\Omega_{k}=\frac{A_{k}}{T}\dfrac{e^{\frac{-(t-T/2)^2}{2(\sigma_kT)^2}}-e^{\frac{-(0-T/2)^2}{2(\sigma_kT)^2}}
	}{1-e^{\frac{-(0-T/2)^2}{2(\sigma_kT)^2}}}e^{\i\theta_k}
\end{align}
The parameters to be optimized include \( A_k \), \( \sigma_k \), and \( \theta_k \), where \( A_k \) denotes the peak amplitude, \( \sigma_k \) characterizes the pulse width, and \( \theta_k \) represents the phase angle.

Taking the three-dimensional Bell state as the target state,
\begin{align}
\ket{\varphi} = \frac{1}{\sqrt{3}} \left( \ket{00} + \ket{11} + \ket{22} \right),
\end{align}
the fidelity is defined as
\begin{align}
f(t) = \bra{\varphi} \rho_{e}(t) \ket{\varphi},
\end{align}
here, \( \rho_e(t) \) denotes the density matrix corresponding to the effective Hamiltonian in Eq. (\ref{belleff}), in the absence of dissipation.
Its evolution follows the quantum Liouville–von Neumann equation
\begin{align}
	\dot{\rho_e}(t)=-\i [H_{\rm eff}(t),\rho_e(t)]
\end{align}
{Our ultimate goal is to minimize the infidelity}, therefore, the objective function can be written as
\begin{align}
	\underset{\{A_k, \sigma_k, \theta_k\}}{\rm Min}~
	\mathlarger{\mathlarger{\{}} 1 - f(T) \mathlarger{\mathlarger{\}}}
\end{align}
To efficiently determine the appropriate Rabi frequencies, we divide the optimization process into two steps.

In the first step, we assume that all Rabi frequencies are real-valued, while the state evolution still occurs in the complex Hilbert space. Therefore, we focus solely on population dynamics in this stage.
In the second step, we introduce phases to the Rabi frequencies and optimize only the parameters $\theta_k$. It is important to note that the parameters \{$A_k$, $\sigma_k$, $\theta_k$\} are not unique; different strategies or variations within the same strategy may lead to distinct solutions.
For optimization, we employ a gradient-based algorithm in the first step, while a genetic algorithm is utilized in the second step. Ultimately, we obtain a set of optimized parameters, as shown in Table \ref{Pabell}.

\begin{table}[ht]
	\centering
	\renewcommand{\arraystretch}{1.4}
	\begin{tabular}{c@{\hskip 50pt}c@{\hskip 50pt}c@{\hskip 50pt}c}
		\toprule
		\boldmath{$k$} & \boldmath$A_k$ & \boldmath$\sigma_k$ & \boldmath$\theta_k$ \\
		\midrule
		1 & 2.94874 & 0.24857 & 0.00000 \\
		2 & 6.47073 & 0.24714 & 3.14148 \\
		3 & 5.64579 & 0.24667 & -0.76998 \\
		4 & 9.28367 & 0.25359 & -2.37169 \\
		\bottomrule
	\end{tabular}
	\caption{{Each pair of optimized pulse parameters for generating a 3D Bell state.}}\label{Pabell}
\end{table}

\subsection{Three-dimensiona GHZ state}
\begin{figure}[ht]
    \centering
    \includegraphics[width=0.5\linewidth]{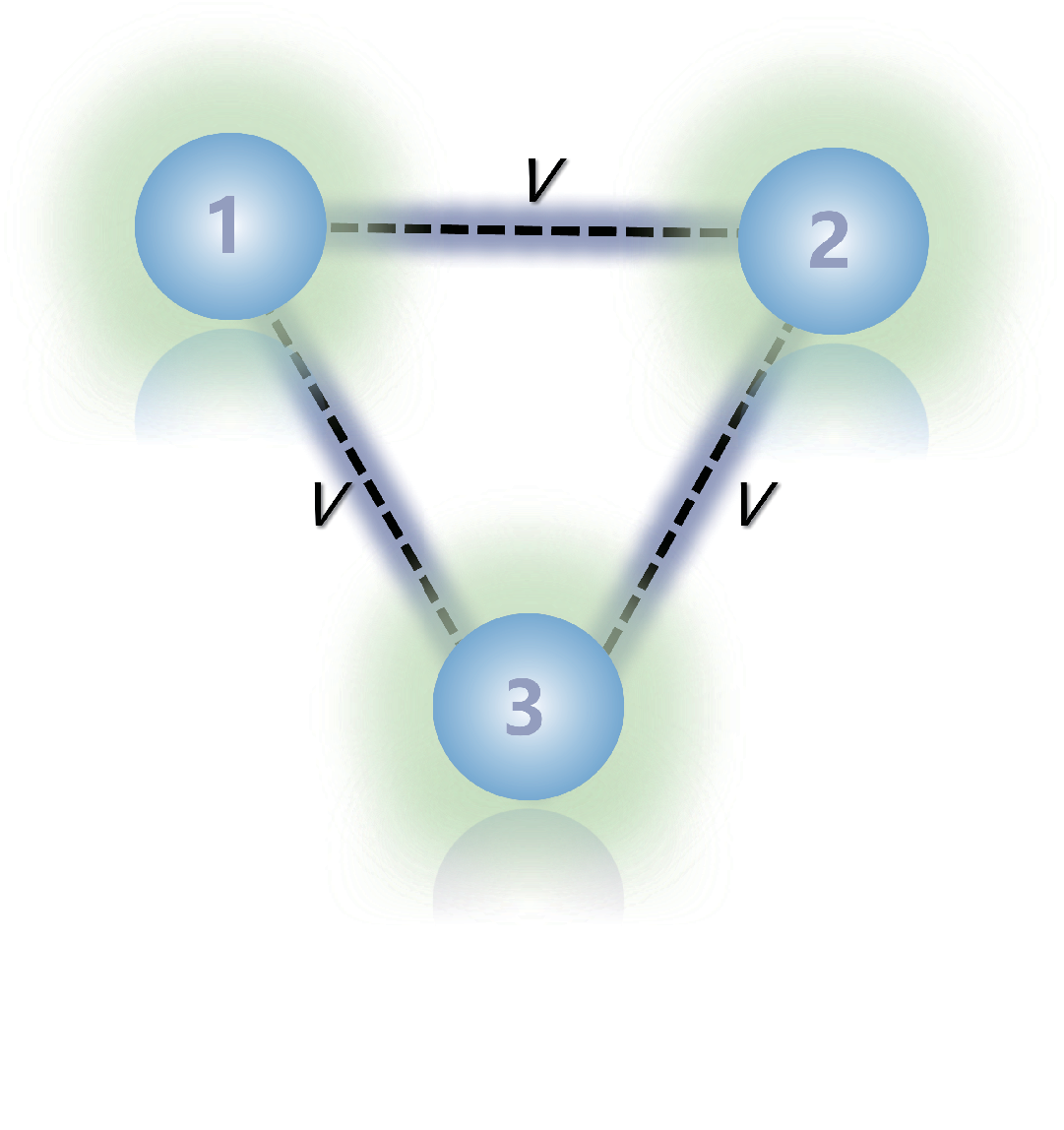}
    \caption{Three identical neutral atoms arranged in a triangular configuration.}\label{三原子}
\end{figure}
We consider three same neutral atoms arranged in a triangular configuration as show in Figure \ref{三原子}. 
The Hamiltonian of a three-level system can be written as
\begin{align}
    H_3(t)=\sum_{k=1,2,3}\sum_{j=0,1}\left[\Omega_{kj}(t)e^{-\i \delta_{kj}t}\ket{2}_k\bra{j}+{\rm H.c.}\right]+\sum_{m,n,m>n}V\ket{2}_m\bra{2}\otimes \ket{2}_n\bra{2}
\end{align}

When setting $\delta_{10}=\delta_{31}=0$, $\delta_{20}=\delta_{11}=V$, $\delta_{30}=\delta_{21}=2V$.
Therefore, when this scheme is extended to a three-atom system, the evolution space naturally includes the basis states$\ket{000}$, $\ket{111}$ and $\ket{222}$, {similar Eq}.(\ref{eq1}) to Eq.(\ref{belleff}), {we can obtain}
\begin{align}\label{A8}
	{H}_3'=&\Omega_{10}\ket{200}\bra{000}+\Omega_{20}\ket{220}\bra{200}+\Omega_{11}\ket{220}\bra{120}+\Omega_{30}\ket{222}\bra{220}\cr&+\Omega_{21}\ket{222}\bra{212}+\Omega_{11}\ket{212}\bra{112}+\Omega_{21}\ket{112}\bra{111}
    +\Hc
\end{align}

In this case, the effective energy-level structure forms an eight-level system. To further simplify the dynamical process, we introduce detuning or waveform control
\begin{align}
     \Omega_{20}\rightarrow \Omega_{20}\cos{\omega t}\cr
     \Omega_{30}\rightarrow \Omega_{30}\i \sin{\omega t}  
 \end{align}
 enabling the system dynamics to be effectively reduced to a chain-like coupling structure, which facilitates the preparation of the desired entangled state.
Under the condition $V\gg{\omega}\gg{\Omega_{ij}}$, 
the effective Hamiltonian become 
 \begin{align}\label{ghzeff}
     H_{\rm fe}=\Omega_1 \ket{200}\bra{000}+\Omega_2 \ket{222}\bra{200}+\Omega_3 \ket{222}\bra{212}+\Omega_4 \ket{212}\bra{112}+\Omega_5 \ket{112}\bra{111}+\Hc
 \end{align}
 where
 $\Omega_1=\Omega_{10}$,  $\Omega_2=\Omega_{20}\Omega_{30}/2\omega$,
$\Omega_3=\Omega_{21}$,
$\Omega_4=\Omega_{11}$,
$\Omega_5=\Omega_{31}$,
This reduces the system to an effective chain-like six-level structure. Analogous to the pulse coupling design for the Bell state, we obtain a set of optimized parameters, as presented in Table \ref{tableghz}.

\begin{table}[ht]
	\centering
	\renewcommand{\arraystretch}{1.4}
	\begin{tabular}{c@{\hskip 50pt}c@{\hskip 50pt}c@{\hskip 50pt}c}
		\toprule
		\boldmath{$k$} & \boldmath$A_k$ & \boldmath$\sigma_k$ & \boldmath$\theta_k$ \\
		\midrule
		1 & 7.07187 & 0.25219 & 2.02993 \\
		2 & 5.76400 & 0.25323 & 1.11166 \\
		3 & 9.46614 & 0.25667 & -2.47423 \\
		4 & 3.87972 & 0.24955 & 2.66979 \\
		5 & 7.65357 & 0.25311 & -1.76635 \\
		\bottomrule
	\end{tabular}
	\caption{{Each pair of optimized pulse parameters for generating a 3D GHZ state.}}\label{tableghz}
\end{table}

\section{Numerical simulations}\label{sec3}

It is noted that the Bell state components include two Rydberg excitations $\ket{22}$, which introduce a relative phase factor $e^{ -\i Vt}$ under the Hamiltonian (\ref{eq1}), i.e.
\begin{align}
    \ket{\psi}=(\ket{00}+\ket{11}+e^{ -\i Vt}\ket{22})/\sqrt{3}
\end{align}
Therefore, the specific dynamics of the chosen system require precise temporal monitoring to account for this relative phase accumulation,  when necessary, local correction can be performed by applying a single-qubit phase gate to any one of the qubits\cite{pachniak2021creation,zheng2020deterministic,wang2025deterministic,li2018engineering}. Here, we do not explicitly consider this relative phase; accordingly, the fidelity is defined as follows
\begin{align}
    F(t)=\bra{\psi}\rho(t)\ket{\psi}
\end{align}

Next, we perform simulations based on Hamiltonian (\ref{eq1}) to demonstrate the feasibility of our results.
Since completely isolating the physical system from its environment is a challenging task, when considering the interaction between the system and its environment, the evolution of the system under the influence of dephasing and spontaneous emission can be described by the master equation\cite{medina2019pulse}
\begin{align}
\frac{\mathrm{d} \rho(t)}{\mathrm{d} t} = -\i \left[ H(t), \rho(t) \right] + \frac{1}{2} \sum_{k}\sum_{\substack{\alpha=0,1 \\ \beta=\gamma,\Gamma}} \left\{ 2 L_{\alpha k}^\beta \rho(t) L_{\alpha k}^{\beta\dagger} - \left[ L_{\alpha k}^{\beta\dagger} L_{\alpha k}^\beta \rho(t) + \rho(t) L_{\alpha k}^{\beta \dagger} L_{\alpha k}^\beta \right] \right\}
\end{align}
$L_{\alpha k}^\beta$are the Lindblad operators used to describe decoherence processes, Here $k$ denotes the $k$th atom, $\beta$refers to the decoherence channels including atomic spontaneous emission and dephasing. Specifically, atomic spontaneous emission is represented by $L_{\alpha k}^\gamma=\sqrt{\gamma/2}\ket{\alpha}_k\bra{2}$, 
and dephasing is represented by $L_{\alpha k}^\Gamma=\sqrt{\Gamma/2}(\ket{2}_k\bra{2}-\ket{\alpha}_k\bra{\alpha})$.
For convenience, we have assumed that each atom decays from $\ket{2}$ to $\ket{0}$ and $\ket{1}$at equal rates denoted by $\gamma$ and that each dephasing rate is uniform, denoted by $\Gamma$.

\begin{figure}[h]
	\centering
	\includegraphics[width=0.5\linewidth]{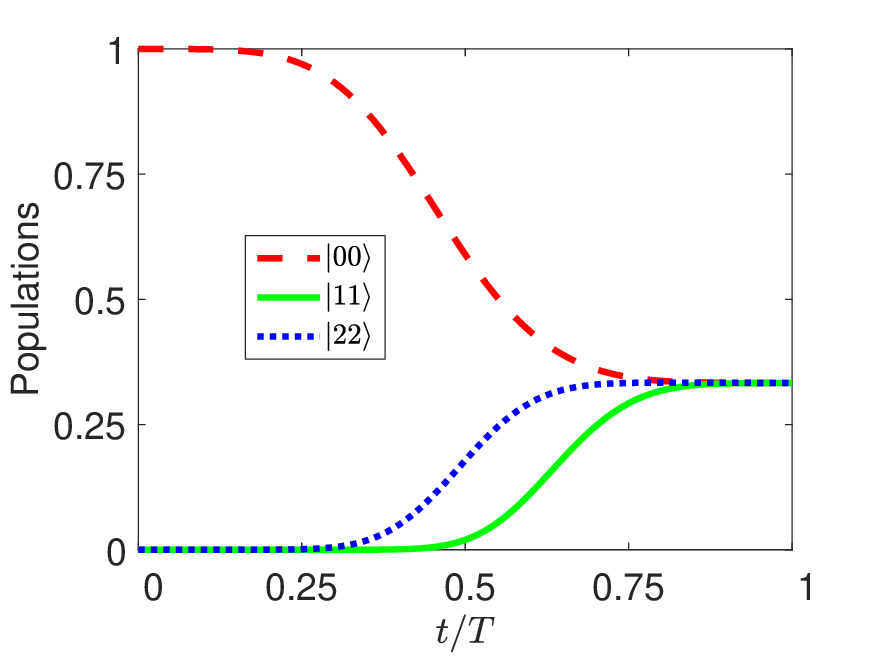}
	\caption{Time evolution of the populations for the states $\ket{00}$, $\ket{11}$, $\ket{22}$.}
	\label{bellpop}
\end{figure}

\begin{figure}[h]
    \centering
  \includegraphics[width=0.5\linewidth]{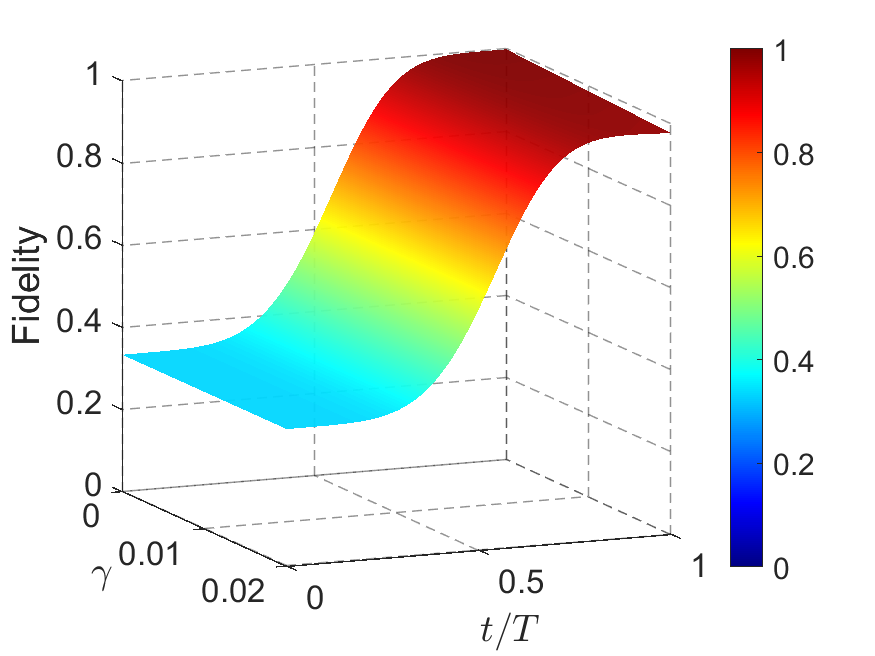}
    \caption{The fidelity of the 3D Bell states versus the interaction time.}
    \label{Bell-F1}
\end{figure}

\begin{figure}
    \centering
    \includegraphics[width=0.5\linewidth]{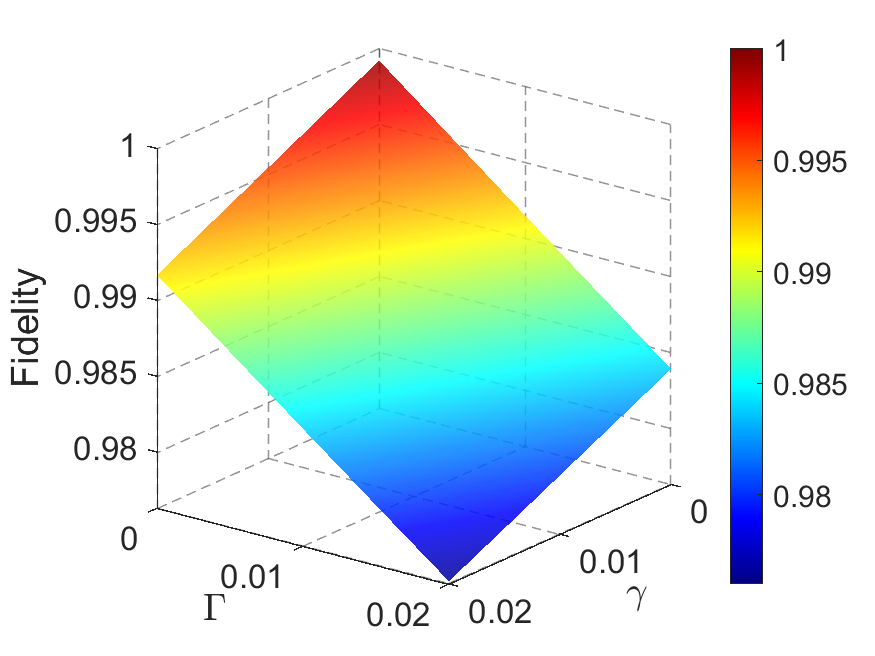}
    \caption{The fidelity versus the decoherence of dephasing $\gamma$ and spontaneous emission  $\Gamma$ at evolution time  $T$($\gamma$,  $\Gamma$ unites of $1/T$).} 
    \label{Bell-F2}
\end{figure}

Here, the condition $V \gg {\Omega}$ must be satisfied. Since  ${\rm Max}\{\Omega\}\lesssim 10/T $, we simply choose $V=200/T$, and  we plot the time evolution of the populations in Figure \ref{bellpop}.
In Figure \ref{Bell-F1}, we plot the fidelity versus the decoherence $\gamma$ and the evolution time $t/T$ .
As shown in Figure \ref{Bell-F2}, the proposed scheme is more sensitive to dephasing than to spontaneous emission. Nevertheless, overall, the scheme demonstrates strong robustness against decoherence effects caused by both atomic spontaneous emission and dephasing. Even when 
 $\gamma=\Gamma=0.02$, the final fidelity remains above 97.6\%.

\begin{figure}[h]
    \centering
    \includegraphics[width=0.5\linewidth]{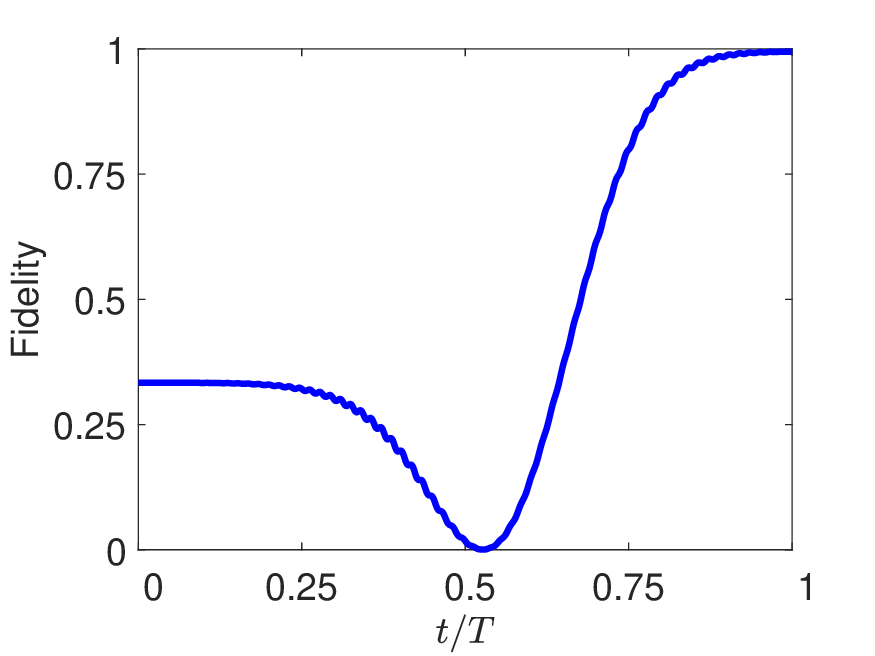}
    \caption{The fidelity of the 3D GHZ states versus the interaction time.} \label{FGHZ}
\end{figure}

\begin{figure}
    \centering
    \includegraphics[width=0.5\linewidth]{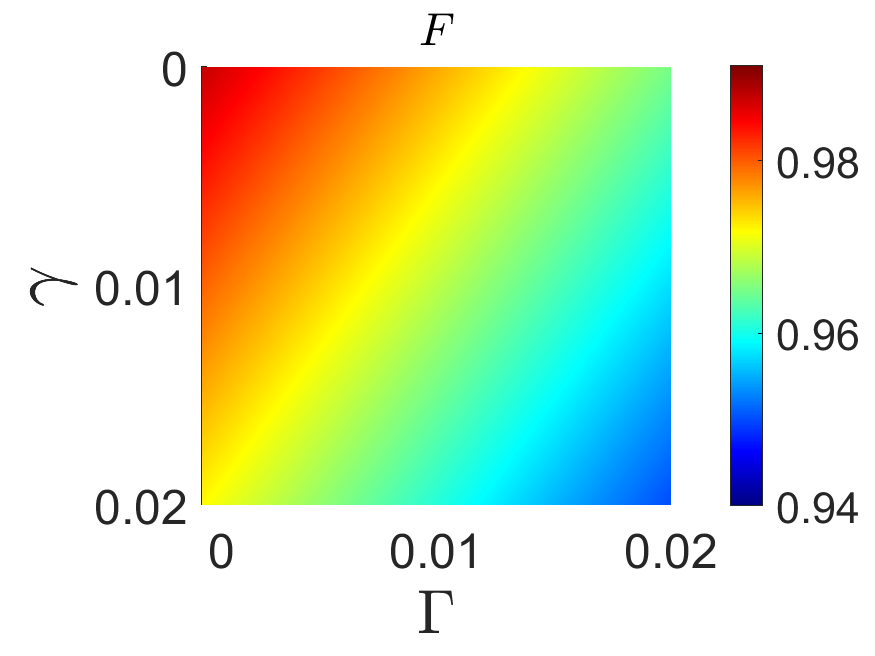}
    \caption{The fidelity versus the decoherence of dephasing $\gamma$ and spontaneous emission  $\Gamma$ at evolution time  $T$($\gamma$,  $\Gamma$ unites of $1/T$).} \label{FGHZ2}
\end{figure}

For the effective pulses related to the GHZ state, we utilize the data presented in Table \ref{tableghz}. The values of $\Omega_{20}$, $\Omega_{30}$ can be determined inversely; for simplicity, we set $\Omega_{20}$=$\Omega_{30}$=$\sqrt{2\omega\Omega_2}$, which maintains a smooth profile. Similarly to the Bell state, this three-dimensional GHZ state also possesses a relative phase
\begin{align}
	\ket{\psi}=(\ket{000}+\ket{111}+e^{ -\i 3Vt}\ket{222})/\sqrt{3}
\end{align}
and we plot the fidelity evolution over time in Figure \ref{FGHZ}, and the impact of spontaneous emission and dephasing on the fidelity is shown in Figure \ref{FGHZ2}, Here, the condition $V\gg \omega\gg \{\Omega\}$ must be satisfied. We set $\omega=200/T$ and  $V=20\omega$. As shown in Figure \ref{FGHZ}, the fidelity initially decreases and then increases. This behavior is attributed to interference effects and the specific design of the evolution paths, indicating that the preparation process is not monotonically optimized. Future work could potentially improve this by introducing additional strategies, such as modifying the pulse shapes or incorporating monotonicity rewards into the optimization. Moreover, we observe high-frequency oscillations in the middle of the fidelity curve. This is due to the fact that the chosen parameter regime does not fully satisfy the "much greater than" hierarchy assumed in the approximations. As shown in Figure \ref{FGHZ2}, under these parameter settings, the fidelity of the GHZ state is lower than that of the Bell state but still remains above 95.6\%.


\begin{figure}
    \centering
    \includegraphics[width=0.9\linewidth]{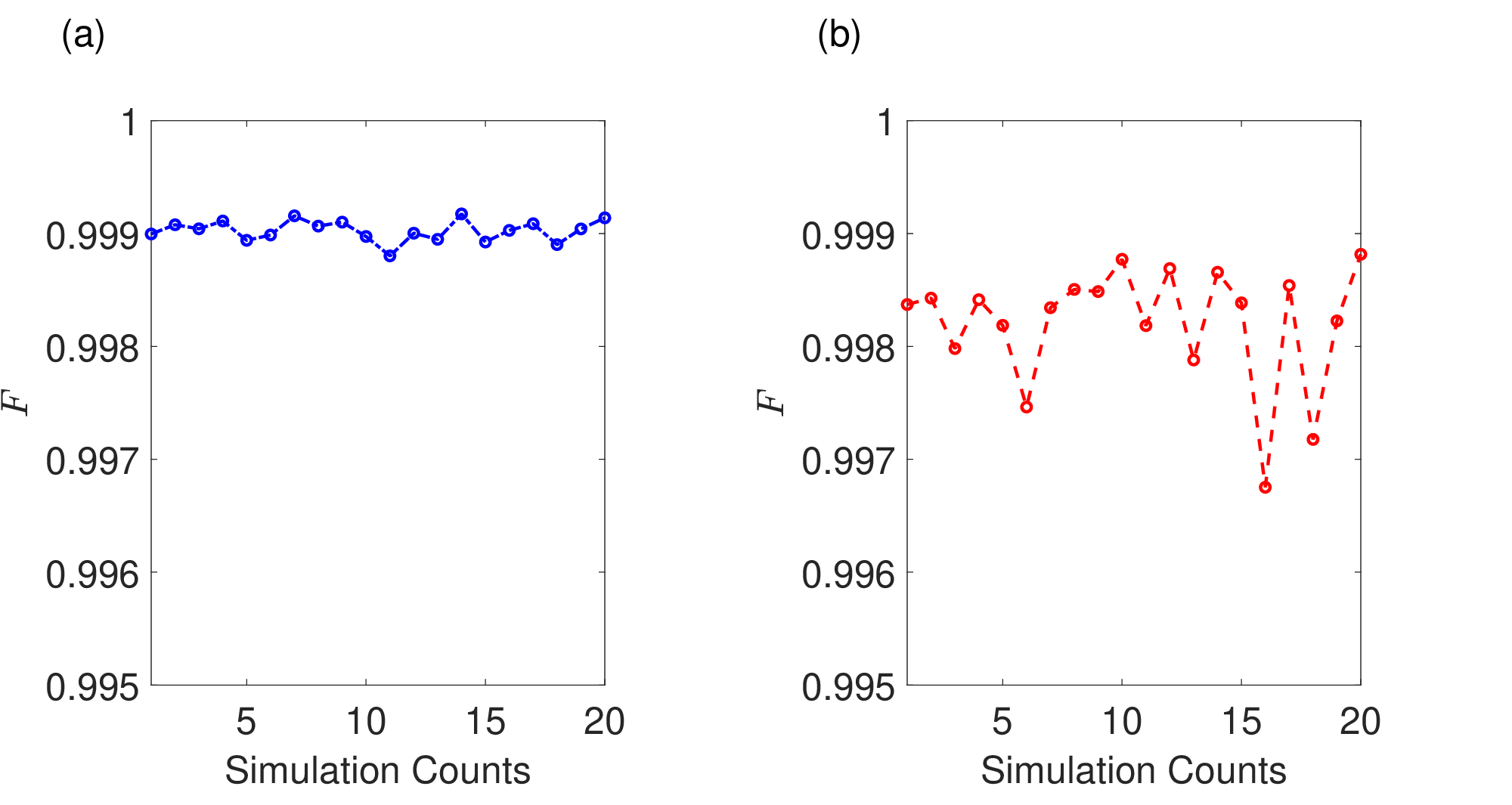}
    \caption{The fidelity of the 3D Bell state versus simulation counts with the random noise of pulse amplitudes. (a) $\mathcal{R}=0.05$ and  (b) $\mathcal{R}=0.1$.} 
    \label{bellrandnoise}
\end{figure}

\begin{figure}
    \centering
    \includegraphics[width=0.9\linewidth]{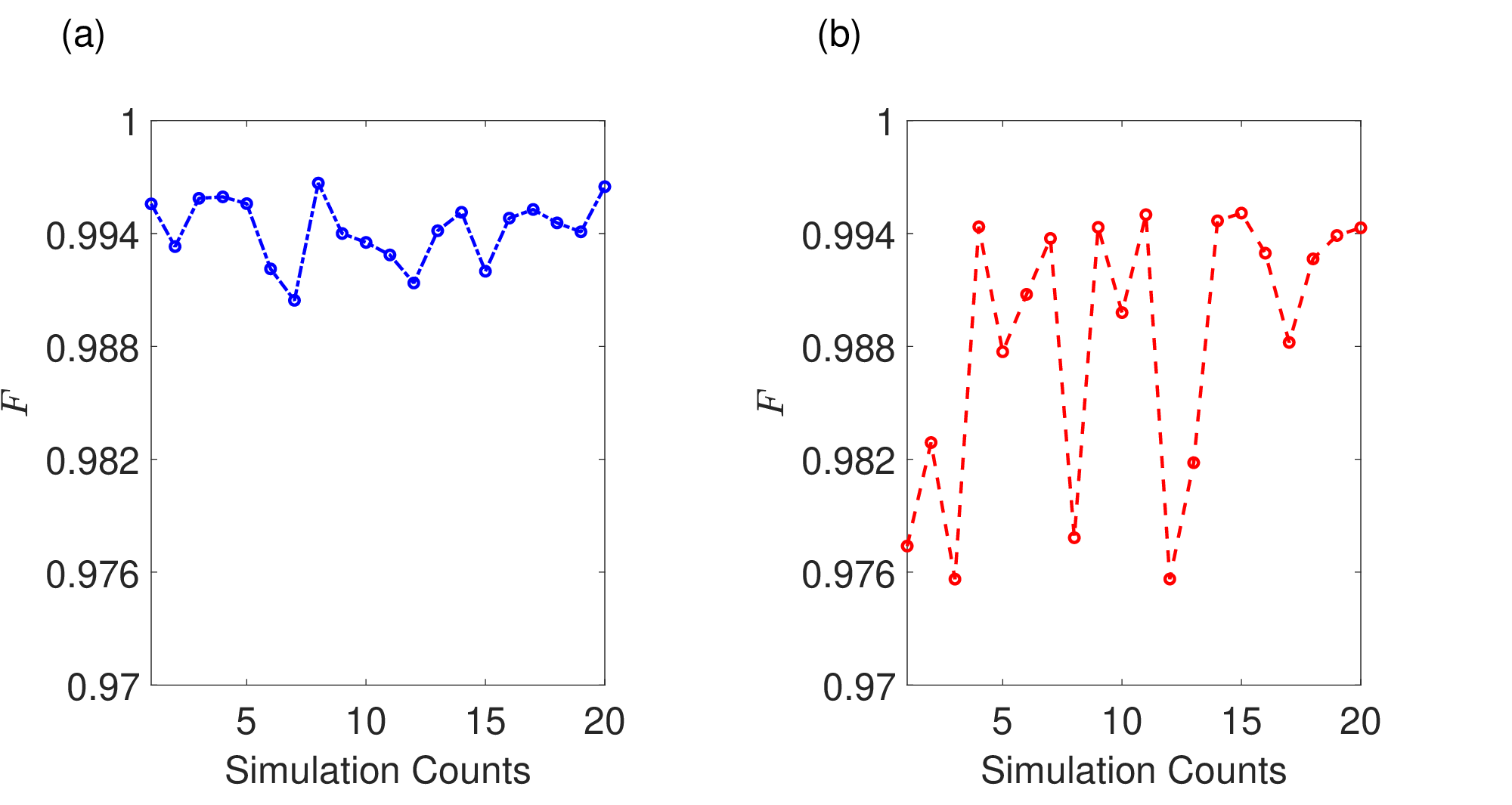}
    \caption{The fidelity of the 3D GHZ state versus simulation counts with the random noise of pulse amplitudes. (a) $\mathcal{R}=0.05$ and  (b) $\mathcal{R}=0.1$.} 
    \label{ghzrandnoise}
\end{figure}

{Moreover, noise can distort the waveform of the Rabi frequency within the pulses during experiments, making robustness against noise a crucial factor in evaluating the feasibility of the scheme. Therefore we introduce random noise independently to each Rabi frequency in the driving pulses, such that the noisy Rabi frequency for each transition takes the form}
\begin{align}
    \Omega^\mathcal{R}(t) = \bigl(1+\text{rand}(\mathcal{R},t)\bigr)\Omega(t),
\end{align}
{where $\text{rand}(\mathcal{R},t)$ generates different random numbers within the range $(-\mathcal{R},\mathcal{R})$ at each time step. We consider two  noise amplitudes, $\mathcal{R}=0.05$ and $\mathcal{R}=0.1$ , since the random noise varies across realizations, we perform 20 independent simulations for each noise amplitude and plot the corresponding fidelities of the generated Bell and GHZ states as show in Figure} \ref{bellrandnoise} and Figure \ref{ghzrandnoise}, {respectively. The results demonstrate that the proposed scheme exhibits considerable robustness against noise.}

In the experiment, we may employ $^{87}\rm Rb$ atoms in this scheme\cite{archimi2019measurements,mack2015all,wangdeterministic,zheng2021fast,yang2024fast,shi2019fast,wu2021one,wang2017single,li2019optical}, {encoding states $\ket{0}$ and $\ket{1}$ in the hyperfine ground states $\ket{5S_{1/2},F = 1}$ and $\ket{5S_{1/2},F = 2}$, respectively. And the Rydberg state is encoded in $\ket{2} = \ket{75P_{3/2}}$}.
When consider a reported parameter the total interaction time of the 3D Bell  state generation is $T = 10\mu s$, which is still much less than the reported coherence time of Rydberg atoms. And the spontaneous emission rate of the Rydberg atoms is $\gamma \sim 2 {\rm kHz}$ and dephasing rate $\Gamma \sim 1 {\rm kHz}$. For Bell state generation, with an interaction strength $V\sim 20{\rm MHz}$ the fidelity can reach 98.4\%.
For GHZ state generation, with $\omega\sim 20{\rm MHz}$ and $V\sim 400{\rm MHz}$ the fidelity can reach 96.6\%.
\section{Summary}\label{sec4}
In this work, the optimization of pulse parameters is solely aimed at achieving a final fidelity close to unity. In the future, additional strategies could be incorporated for further improvement, such as minimizing the population of non-target states and adaptively tuning pulse shapes. Moreover, while this study first designs pulses for the effective Hamiltonian and then verifies them with the original Hamiltonian, it is also possible to directly optimize and match the parameters of the original Hamiltonian in future research.
We hope that this scheme can provide valuable help in future quantum information processing
tasks.

\section*{Acknowledgments}{
	We  acknowledge
	Li-Ping Yang for comments on various stages
	of the manuscript.}

\providecommand{\newblock}{}
\providecommand{\url}[1]{{\tt #1}}
\providecommand{\urlprefix}{}
\providecommand{\href}[2]{#2}

\end{document}